\documentstyle[12pt,epsfig]{article}
\def\gappeq{\mathrel{\rlap {\raise.5ex\hbox{$>$}}
{\lower.5ex\hbox{$\sim$}}}}

\def\lappeq{\mathrel{\rlap{\raise.5ex\hbox{$<$}}
{\lower.5ex\hbox{$\sim$}}}}

\def\beq{\begin{equation}}
\def\eeq{\end{equation}}
\def\bea{\begin{eqnarray}}
\def\eea{\end{eqnarray}}

\def\Toprel#1\over#2{\mathrel{\mathop{#2}\limits^{#1}}}
\def\Pbar{\Toprel{\lower 2pt
\hbox{$\scriptscriptstyle(-)$}}\over{\partial}}

\def\Toprel#1\over#2{\mathrel{\mathop{#2}\limits^{#1}}}
\def\Gbar{\Toprel{\lower 2pt
\hbox{$\scriptscriptstyle(-)$}}\over{G}}

\def\Toprel#1\over#2{\mathrel{\mathop{#2}\limits^{#1}}}
\def\pbar{\Toprel{\lower 2pt
\hbox{$\scriptscriptstyle(-)$}}\over{P}}

\def\Toprel#1\over#2{\mathrel{\mathop{#2}\limits^{#1}}}
\def\Pibar{\Toprel{\lower 2pt
\hbox{$\scriptscriptstyle(-)$}}\over{\Pi}}

\begin{document}
\begin{titlepage}
\pagestyle{empty}
\begin{flushright}
{CERN-TH.7137/94}
\end{flushright}
\vspace*{1mm}
\begin{center}
{\bf THE VALUES OF $m_t$ AND $\bar{\alpha_s}$ DERIVED FROM THE
NON-OBSERVATION}\ {\bf OF ELECTROWEAK RADIATIVE CORRECTIONS AT
LEP: GLOBAL FIT} \\
\vspace{0.2in}
{\bf  V.A. Novikov}$^{*)}$ ,
{\bf L.B.  Okun}$^{*)}$ \\
\vspace{0.1in}
Theoretical Physics Division, CERN\\
CH-1211 Geneva 23, Switzerland \\
\vspace{0.1in}
{\bf A.N. Rozanov}$^{**)}$ \\
\vspace{0.1in}
Particle Physics Experiments Division, CERN\\
CH-1211  Geneva 23, Switzerland \\
and \\
\vspace{0.1in}
{\bf M.I. Vysotsky, V.P. Yurov}\\
\vspace{0.1in}
ITEP, Moscow 117259, Russia\\
\vspace{0.3in}
 {\bf Abstract \\}
 \end{center}
 A set of equations
     representing the $W/Z$ mass ratio and various observables of
 $Z$ decays in terms of $\bar\alpha \equiv\alpha (m_Z)$, $G_{\mu}$, $m_Z$,
 $m_t$, $m_H$, $\bar\alpha_{s} \equiv\alpha_{s} (m_Z)$, $m_b$ and $m_\tau$
(all other fermion masses being neglected) are compared with the latest
data of the four LEP detectors, which at the level of one standard deviation
coincide with their Born values.
Our global fit gives:  $m_t = 161^{+ 15
+16}_{-16 -22} \;\; , \;\; \bar\alpha_{s} = 0.119 \pm 0.006 \pm 0.002$,
where the central values correspond to $m_H = 300$ GeV, the first errors are
statistical and the second ones represent shifts of the central values
corresponding to $m_H = 1000$ GeV($+$) and 60 GeV($-$). The
predicted mass of the top is smaller than in the recent
fits by 4 GeV.
  The {\it predicted} values of $m_W/m_Z$ and the LEP observables,
based on the fitted values of $m_t$ and $\bar{\alpha}_s$, show a weak
dependence on $m_H$ and differ by several  {\it predicted} standard
deviations from the corresponding Born values. The uncertainties of the
predicted values and their deviations  from the  corresponding Born values
determine the experimental accuracy required to observe electroweak
radiative corrections.

\vspace{0.1in}
\noindent
\rule[.1in]{16.5cm}{.002in}
\noindent
$^{\; *)}$ Permanent address: ITEP, Moscow 117259, Russia. \\
$^{**)}$ On leave of absence from ITEP, Moscow 117259, Russia.

\begin{flushleft}
CERN-TH.7137/94 \\
January 1994
\end{flushleft}
\end{titlepage}

\vfill\eject
\pagestyle{empty}

\setcounter{page}{1}
\pagestyle{plain}

Precision measurements of 5 million events of $Z$ decays observed at LEP has
 allowed a test of the Standard Model of
 unprecedented accuracy. It turned out
 \cite{aaa} that
   all these data, for each observable
 separately,
 can be described
  within $ 1 \sigma$
  by
 a simple Born approximation, using as a basis the value of the running
electromagnetic $\alpha$ at $q^2 = m^{2}_{Z}$:
\begin{equation}
\bar{\alpha} \equiv
\alpha(m_Z^2) = 1/128.87(10) \;\; ,
\label{1}
\end{equation}
the Fermi coupling constant:
\begin{equation}
G_{\mu} = 1.16639(2)\times 10^{-5} \, \mbox {\rm GeV}^{-2} \;\; ,
\label{2}
\end{equation}
and the mass of the $Z$ boson:
\begin{equation}
m_Z = 91.187(7) \, \mbox{\rm GeV} \;\; .
\label{3}
\end{equation}

The non-observation of genuinely electroweak radiative corrections by the
high precision LEP experiments
allows us within the framework of the Standard Model, to obtain
rather accurate
predictions of the mass of the top quark and of the gluon coupling constant
at $q^2 = m^2_Z$, $\bar\alpha_s = \bar\alpha_s(m_Z^2)$. In this paper we
present the results of a global fit of all LEP data and $m_W/m_Z$ ratio
within the Standard Model one-loop approximation based on three most
accurately measured observables: $G_\mu$, $m_Z$ and $\bar\alpha$.  Our
results for $m_t$ and $\bar{\alpha}_s$ are close to but slightly different
from those obtained by other authors, who based their global fits on other
parametrizations in which the running of $\alpha$ was not separated from
genuinely electroweak corrections. In particular, our central value of $m_t$
is 2--4 GeV lower than in the previous fits.

The predicted theoretical values of $m_W/m_Z$ and of various LEP observables
(including their uncertainties) calculated with the fitted values of
$m_t\pm\delta m_t$ and $\bar{\alpha}_s\pm\delta\bar{\alpha}_s$ are compiled
in Table 1. It can be seen that the predicted values differ by several
standard deviations from their $\bar{\alpha}$ Born values. Table 1 shows
what accuracy should be reached by experiments in order to observe the
electroweak radiative corrections and to test in this way the Standard
Model.

Four relations describing four ``gluon-free'' observables
$m_W/m_Z$, $g_{\nu}$, $g_A$ and $g_V/g_A$ in terms of
$\bar\alpha \equiv\alpha (m_Z^2)$,
 $G_{\mu}$ and $m_Z$,
and of the
masses of the top quark $m_t$ and Higgs boson $m_H$ have been derived
in ref. \cite{bb} in electroweak one-loop approximation:
\begin{equation}
m_W /m_Z = c+\bar{\alpha}\frac{3c}{32\pi s^2(c^2 -s^2)}V_m(t,h) \;\; ,
\label{4}
\end{equation}
\begin{equation}
g_\nu = \frac{1}{2}+\bar{\alpha}\frac{3}{64\pi s^2 c^2}
V_\nu(t,h) \;\; ,
\label{5}
\end{equation}
\begin{equation}
g_A = -\frac{1}{2}-\bar{\alpha}\frac{3}{64\pi s^2 c^2}
V_A(t,h) \;\; ,
\label{6}
\end{equation}
\begin{eqnarray}
g_V/g_A &=& 1-4s^2 +\bar{\alpha}\frac{3}{4\pi(c^2- s^2)}
V_R(t,h) \;\; ,
\label{7}
\end{eqnarray}
where $t=(m_t/m_Z)^2$, $h=(m_H/m_Z)^2$, while
$c\equiv \cos\theta$ and $s\equiv\sin\theta$ were defined by
\begin{equation}
 c^2 s^2 =
\frac{\pi\bar{\alpha}}{\sqrt{2}G_{\mu}m^2_Z} \;\; .
\label{8}
\end{equation}

{}From Eqs. (\ref{1}), (\ref{2}), (\ref{3}), (\ref{8}) it follows that
\begin{equation}
s^2 = 0.23118(33) \;\; .
\label{9}
\end{equation}

Explicit expressions of $V_m$, $V_\nu$, $V_A$, $V_R$ in terms of
$t$ and $h$
and parameters $s$ and $c$ are given by eqs. (40), (68), (78) and (103)
of ref. \cite{bb}.

Each function $V_i$ is a sum of four terms:
\begin{equation}
V_i(t,h)=t+T_i(t)+H_i(h)+C_i \;\; .
\label{10}
\end{equation}

Explicit expressions for $T_i(t)$ in terms of $t, c, s$ are given in ref.
 \cite{bb}: eq.(42) for $T_m$; eq. (84) for $T_A$; eq. (104) for $T_R$.

As for the $H_i(h)$ they have been presented in \cite{bb} by explicit
functions of $h, c, s$ plus certain, rather cumbersome, combinations of
 $c$ and $s$ that were not written out but were evaluated in \cite{bb}
numerically for $s^2 = 0.2315 (3) $ and $m_Z = 91.175$ GeV.  For the updated
values of the above combinations and constants $C_i$ in eq. (\ref{10}), see
Appendix B of ref. \cite{cc}.

In ref. \cite{dd} the contribution of virtual gluons in the quark
 loops have been taken into account;
for the sum of light and heavy quark loops
up to terms $O(\frac{1}{t^3})$ [misprints of the preprint \cite{dd} has been
corrected in the text to be published in Yadernaya Fizika \cite{dd}]:
\begin{eqnarray}
\delta V_m(t) &=& \frac{\bar\alpha_s}{\pi} \left( -2.86 t + 0.46 \ln t -1.92
- \frac{0.68}{t} - \frac{0.21}{t^2} \right) \;\; ,
\label{11}\\
\delta V_A(t) &=& \frac{\bar\alpha_s}{\pi} \left( -2.86 t + 2.24
- \frac{0.19}{t} - \frac{0.05}{t^2} \right) \;\; ,
\label{12}\\
\delta V_R(t) &=& \frac{\bar\alpha_s}{\pi} \left( -2.86 t + 0.22 \ln t -1.51
- \frac{0.42}{t} - \frac{0.08}{t^2} \right) \;\; .
\label{13}
\end{eqnarray}
The expressions (\ref{11})-(\ref{13}) are valid for $t \geq 1$.
For $m_t < m_Z$ we either put
 $\delta V_m = 0 \, , \; \delta V_A = 0 \, , \;\delta V_R = 0$,
 or used  interpolation to the massless limit in which
\begin{equation}
\delta V_m = \frac{\bar{\alpha}_s}{\pi}4(c^2-s^2)\ln c^2 =
-\frac{\bar{\alpha}_s}{\pi} \cdot 0.57
\label{a}
\end{equation}
\begin{equation}
\delta V_A = \frac{\bar{\alpha}_s}{\pi}4\left(c^2-s^2+ \frac{20}{9}s^4\right)
= +\frac{\bar{\alpha}_s}{\pi} \cdot 2.63
\label{b}
\end{equation}
\begin{equation}
\delta V_R = 0.
\label{c}
\end{equation}

As the corrections are anyway small at $t<1$ they practically do not change
the fitted values of $m_t$ and $\bar{\alpha}_s$.
In the global fit condition,
we also used
$t\geq 1$, which is definitely valid in view of the recent
results \cite{104} of CDF and D0 ($m_t > 108$
GeV at 95\% CL).  This latter procedure does not change the fitted values
of $m_t$ and $\bar{\alpha}_s$.

In ref.\, \cite{ee} the same approach based on the
 $\bar{\alpha}$, $G_{\mu}$, $m_Z$
parametrization and on the functions $V_i$  was used to
calculate the hadronic $(q \bar{q})$ decays of $Z$. In the gluonless
approximation the differences of $V$'s for quarks and leptons are given by
eqs. (7)--(10) of ref. \cite{ee}.
In this approximation the largest two-loop terms should be considered, which
are proportional to $m_t^4$: one of them is in the $t\bar{t}$ contribution
to the $Z$ boson self energy, the second --  in the $t\bar{t}$ contribution
to $Z\to b\bar{b}$ vertex. These terms have been calculated in ref.
\cite{07}. They become noticeable for large values of $m_H/m_t$. To take
into account the first of them one has to multiply the term $t$ in eq.
(\ref{10}) of the present paper by factors in brackets derived from
equations  (16a) and (17a) of ref. \cite{07}.  To take into account the
second term one has to multiply the term $t$ in the eq. (18) of ref.
\cite{ee} by factors in brackets from equation (16b) and (17b) of ref.
\cite{07}. Note that Eqs. (16) refer to $m_H = 60$ GeV, while Eqs. (17) --
to $m_H = 300$ and $1000$ GeV.

In order to take into account the virtual gluons in the $Z\to bb$ vertex one
has also to multiply the term $t$ in eq. (18) of ref. \cite{ee} by a factor
($1 - 2.29\alpha_s/\pi$) calculated in refs. \cite{08}, \cite{09}. All other
gluonic corrections, up to $(\alpha_s/\pi)^3$, known in the literature, are
included in the eqs. of ref. \cite{ee}. In particular the
running mass of the $b$-quark was also included in ref. \cite{ee}.
We present the results of our fit for $m_b(m_Z) = 3.1$ GeV according to refs.
\cite{105}, \cite{106}. If we change $m_b (m_Z)$ from 2.4 to 3.4 GeV,
the predicted value of $R_b$ at $m_H = 300$ GeV in \mbox{Table 1}
 decreases from
0.2164 to 0.2158, the fitted central value of $m_t$ decreases by 0.5 GeV,
while that of $\bar{\alpha}_s$ increases by 0.002. The latter will change
the Born values of $\Gamma_h$, $\Gamma_Z$, $\sigma_h$, $R_l$. All masses of
fermions lighter than $b$ give very small contributions. Thus inclusion of
$m_{\tau}$ decreases $\Gamma_{\tau}$ by 0.19 MeV. As for the running mass of
the charmed quark ($m_c (m_Z)< 1$ GeV), we neglect it.

It should be emphasized that not all two-loop corrections of the order of
$\alpha\alpha_s$ have been calculated in the literature. In particular the
vertex triangle electroweak diagrams with a gluon connecting a quark line in
the triangle with an external quark line are unknown. Such two-loop
corrections may substantially change the values of $m_t$ given in Fig. 3,
but its overall fitted value is expected to be changed by
approximately 1 GeV. This can be seen by using two different expressions for
$\Gamma_b$: one in which the specific corrections due to the $ttW$ triangle and
those due to external gluons are treated separately for the vector and axial
channels (ref. \cite{ee}, eq. (27) \footnote{Note a misprint in
the sign of the third term in the third line of eq. (27) in ref.
\cite{ee}.}), the other in which they enter as a single correction (ref.
\cite{ee}, eq. (17)).

The equations of refs. \cite{aaa}-\cite{dd}, \cite{ee} -- \cite{106}
described above and
the latest experimental LEP data \cite{ff}, \cite{ggg} (see Fig. 3 and Table
1) are used in this paper to fit the values of $m_t$ and $\bar\alpha_s$.
 Assuming $m_H$= 60, 300 and 1000 GeV, the results of the fit are:

\begin{equation}
m_t = 162^{+ 16 +17}_{-17 -23}
\label{14}
\end{equation}
\begin{equation}
\bar\alpha_{s} = 0.119 \pm 0.006 \pm 0.002 \;\; ,
\label{15}
\end{equation}
\begin{equation}
\chi^2 / \mbox{\rm d.o.f.} = 3.5 / 8 \;\; ,
\label{16}
\end{equation}
where the central values correspond to $m_H$ = 300 GeV, the first error is
experimental, the second one corresponds to the variation of $m_H$:
sign $+$ corresponds to $m_H = 1000$ GeV, sign $-$ to $m_H = 60$ GeV.
(Note that according to ref. \cite{010}, for $m_H = 60$ GeV and $m_t$ from
110 to 190 GeV the vacuum is unstable but with a lifetime $> 10^{10}$ yr.;
for $m_t > 190$ GeV the vacuum is dangerously unstable.
For $m_H > 300$ GeV vacuum is stable for any reasonable value of $m_t$.)

Independent constraints on $m_t$ are given by the measurements of
the $m_W/m_Z$ mass ratio on $p \bar{p}$ colliders by UA2 \cite{hh}
and CDF \cite{jj} experiments. The Particle Data Group fit of the
$m_W/m_Z$ ratio from these experiments \cite{kk} is:
\begin{equation}
    m_W / m_Z = 0.8798 \pm 0.0028.
\label{17}
\end{equation}

The combined fit of LEP and $p \bar{p}$ collider data
gives:
\begin{equation}
m_t = 161^{+ 15 +16}_{-16 -22} \;\; ,
\label{18}
\end{equation}
\begin{equation}
\bar\alpha_{s} = 0.119 \pm 0.006 \pm 0.002 \;\; ,
\label{19}
\end{equation}
\begin{equation}
\chi^2 / \mbox{\rm d.o.f.} = 3.5 / 9 \;\; .
\label{20}
\end{equation}

The allowed region of $m_t$ and $\bar{\alpha}_s$ is shown in Fig. 1.
The value of the Higgs mass is fixed here to $m_H$ = 300 GeV.
The lines represent the $s$-standard deviation ellipsoids
 ($s$=1, 2, 3, 4, 5),
corresponding to constant values of $\chi^2$
 ($\chi^2 = \chi^2_{min} + s^2$).
 So in the case of validity of Gaussian errors the projections of the
 ellipsoids on the $m_t$ and $\bar\alpha_s$ axis define $s$-standard
  deviation
confidence intervals of the corresponding parameters.

The existing data do not put real limits on the mass of the Higgs.
The corresponding contour plots are shown in Fig. 2, where $\bar\alpha_s$ was
fixed to $\bar\alpha_s$ = 0.119. The minimum of $\chi^2$ is at a very low
value of $m_H$, but even the two-standard deviation contour is not contained
in the 1000 GeV range of Higgs mass.  Similar results were obtained by many
authors \cite{ff}, \cite{ggg}, \cite {107}-\cite{nn}
by using various computer codes \cite{201}-\cite{209}.
Our fit favours slightly smaller ($\approx$ 4 GeV) values of $m_t$.

The individual values of $m_t$ for each of the LEP observables for
$m_H = 300$GeV and $\bar\alpha_s = 0.119$, are shown in Fig. 3.
The data on  $A^{e}_{\tau}$, $R_l$ and
$\sigma_{had}$ are insensitive to the top mass. The data on
$R_b$ and $R_l$ are
compatible with low values of the $m_t$, excluded by the direct search on
the Tevatron.

According to ref. \cite{oo} there should exist substantial non-perturbative
parts of gluonic corrections, which appear at the $t\bar{t}$ threshold and
enter the $Z$-boson propagator through dispersion relations. These
non-perturbative corrections, according to ref. \cite{oo}, may reach
25--50\% of the perturbative ones. The authors of ref. \cite{155} insist that
these corrections are fully non-controllable and eventually would prevent
extraction of any information about the Higgs from LEP precision measurements.
We consider these results as artifacts of dispersion calculation of Feynman
integrals. When real parts of the same integrals are calculated directly
(without referring to their imaginary parts), it is obvious that
the non-perturbative effects
are absolutely negligible, because of the asymptotic freedom of QCD
and large virtuality of $t\bar{t}$ loop for external momenta not larger than
$m_Z:~ 2m_t -m_Z\geq \, 200 \mbox{\rm GeV} \gg \Lambda_{QCD}$. (See for
instance
ref. \cite{166}, where these arguments were applied to charmonium.) Thus we
neglect $t\bar{t}$ threshold effects in our fit.

However we should note that another source of uncertainty in
$O(\alpha\alpha_s)$ corrections really exists -- the virtuality of quarks in
the vector boson self energy loops varies with varying internal and
external momenta, which influence the value of $\alpha_s$ \cite{dd}.

It is instructive to use the fitted value of $m_t$ in order to predict the
theoretical values of every LEP observable for three values of $m_H = 60 \,
, \; 300$ and $1000$ GeV. The corresponding central values of $m_t$ are,
according to Eq. (\ref{14}), $m_t = 139 \, , \; 162$ and $178$ GeV,
respectively.  The results of this procedure are presented in Table 1. One
can easily see that the central predicted values of observables depend on
$m_H$ rather weakly.  One can also see that the uncertainties of the
predicted values of observables are much smaller than their present
experimental uncertainties; also, the predicted theoretical central values
differ from the corresponding Born values by several standard deviations,
calculated from the fitted errors of $m_t$ and $\bar{\alpha}_s$. For
instance, in the case of $R_b$ it is 4.5$\sigma$--6.5$\sigma$ while in the
case of $m_W/m_Z$ it is $3\sigma$.  The latter fact was used recently by
Sirlin \cite{pp} to argue that in the case of $m_W/m_Z$ the existence of
radiative corrections has been proved at the $3\sigma$ level.  It is
obvious, however, that the high accuracy of the predicted theoretical value
of any of the observables in Table 1 cannot serve as evidence for the
observation of electroweak radiative corrections.  Only direct
high-precision measurements could provide such evidence. At present the
experimental uncertainties accommodate comfortably both the $\bar{\alpha}$
Born values and the central one-loop-corrected values.

The quality of the description of the data is not very different at the tree
level and at the one-loop level and weakly depends on $m_H$. Of course we do
not think that the $\bar{\alpha}$ Born approximation would provide a universal
description of all LEP data when their accuracy would be much
better than now. However at present it seems to be sufficient.

As for the weak dependence on $m_H$, one has to bear in mind that it partly
follows from the correlation between $m_H$ and $m_t$ in the fit: the heavier
is the Higgs the heavier is the top. Therefore when the top is discovered
and its mass is known the limits on the Higgs mass will be more stringent
(see columns 4--7 of the table in ref. \cite{aaa}).

Inspection of Table 1
reveals that before the top is discovered the natural experimental
strategy is to reduce the
experimental uncertainties of the observables to the level of the predicted
ones (by a factor of 4 for $R_b$ and $\sigma_h$, by a factor of 2.5 for
$m_W/m_Z$). After the mass of the top is measured with accuracy $\pm 5$ GeV
the reduction of uncertainties for $g_V/g_A$ by a factor of 4 will be
especially promising. Such a program requires not only increasing the LEP
statistics by a factor of 20 ($10^8$ $Z$ bosons), but -- what may be much
more difficult -- reaching a qualitatively new level of sophistication in
controlling systematics. If this can be done, LEP1 will test the electroweak
corrections of the Standard Model and may possibly reveal new physics.

We do not discuss in this paper other possible manifestations of electroweak
loops, such as in $\nu e$ scattering, deep inelastic $\nu
N$ interaction, parity violation in atoms, $K^0\leftrightarrow
\bar{K}^0$, $B^0\leftrightarrow \bar{B}^0$, $B_s^0\leftrightarrow
\bar{B}_s^0$ transitions, radiative corrections to superallowed
$\beta$ decays, such as $~^{14}$O. In some of these experiments the
accuracy is much worse than in $Z^0$ decays, in others - the
results of calculations heavily depend on nonperturbative QCD effects which
can be estimated only roughly. As for $~^{14}$O the virtual $W$ boson
serves here only to cut off the logarithmically divergent large
electromagnetic correction.  The genuine electroweak corrections must be
isolated from the
trivial electromagnetic ones in order to see whether the former can be
quantitatively compared with available experimental data on $~^{14}$O.

Finally, we would like to present in Table 2 our fit in the traditional form
used by the LEP collaborations in spite of its obvious shortcomings as compared
with Table 1: it contains only two of six independent observables
($s_W^2\equiv\sin^2\theta_W\equiv 1-m^2_W/m^2_Z$ and
$s^2_{eff}\equiv\sin^2\theta_{eff}^{lept}=\frac{1}{4}(1-g_V/g_A)$)
and the way it averages the predicted and measured values of
$s^2_W$
is potentially misleading.

The second and the fourth columns of Table 2 are taken from ref.
\cite{ggg}, the third and the fifth present similar results of our fit
taken from our \mbox{Table 1}.

\newpage

\noindent
{\bf Acknowledgements}

We are grateful to  D.Yu. Bardin, B. Kniehl, A. Sirlin, V. Telegdi and
M. Voloshin for helpful discussions and comments.
Three of us (VAN, LBO, ANR) are grateful
to CERN for warm hospitality.  LBO, VAN, MIV and VPY are grateful to the
Russian Foundation for Fundamental Research for grant 93-02-14431.

\newpage

\begin{center}

{\large Table 1}

\vspace{8mm}

\begin{tabular}{|l|l|l|l|l|}\hline
\hline
Observable & Exp. value & \multicolumn{3}{c|}{Predicted values for
observables, $\bar{\alpha}$ Born values} \\
  &    & \multicolumn{3}{c|}{and radiative corrections.} \\
\cline{3-5}
  &    & $m_H = 60$ GeV & $m_H = 300$ GeV &
$m_H = 1000$ GeV \\ \cline{3-5}
   &   & $m_t =139(17)$ & $m_t = 161(16)$ & $m_t = 178(14)$ \\
   &   & GeV & GeV & GeV \\
\cline{3-5}
   &   & $\bar{\alpha}_s =0.116(6)$ & $\bar{\alpha}_s = 0.119(6)$ &
   $\bar{\alpha}_s = 0.121(6)$ \\ \hline
$m_W/m_Z$ & 0.8798(28) & 0.8796(11) & 0.8799(10) & 0.8799(9) \\
  \cline{3-5} &   & 0.8768(\underline{2})  & 0.8768(\underline{2}) &
0.8768(\underline{2}) \\ \cline{3-5}
  &   & 0.0028(11) & 0.0030(10) & 0.0031(10) \\ \hline
$g_A$ & --0.50093(82) & --0.50067(37) & --0.50092(33) &
-0.50096(30) \\ \cline{3-5}
 &  & --0.50000(0)  &--0.50000(0)  & --0.50000(0) \\ \cline{3-5}
   &  & --0.00067(\underline{32}) & --0.00092(\underline{33}) &
   --0.000096(\underline{32}) \\ \hline
$g_V/g_A$ & 0.0716(28) & 0.0711(21) & 0.0704(19)  & 0.0697(17) \\
\cline{3-5}
  &   & 0.0753(\underline{12})& 0.0753(\underline{12}) &
0.0753(\underline{12}) \\ \cline{3-5}
 &   &--0.0042(19) & --0.0049(19) & --0.0056(18)  \\ \hline
 $\Gamma_l$
(GeV) & 0.08382(27) & 0.08373(15) & 0.08381(13) & 0.08382(12) \\ \cline{3-5}
&   & 0.08357(\underline{2})& 0.08357(\underline{2}) &
 0.08357(\underline{2}) \\ \cline{3-5}
 & & 0.00017(13)& 0.00024(13)  & 0.00025(13)\\ \hline
 $\Gamma_h$ (GeV)
&1.7403(59) & 1.7374(40) & 1.7384(39) & 1.7382(38) \\ \cline{3-5}
&  &
1.7394(36) & 1.7407(36) & 1.7420(36) \\ \cline{3-5}
&   &--0.0020(23)   & --0.0023(24)   & --0.0038(22)  \\
\hline
$\Gamma_Z$ (GeV) & 2.4890(70) &2.4888(48) & 2.4905(45) &2.4904(43)
 \\ \cline{3-5}
 &  & 2.4877(36) & 2.4890(36) & 2.4903(36) \\ \cline{3-5}
  & &0.0011(33) & 0.0015(34)  & 0.0001(33)  \\ \hline
  $\sigma_h$ (nb) & 41.560(140) & 41.463(34) &
 41.466(34) & 41.468(34) \\ \cline{3-5}
 &  & 41.462(\underline{34}) & 41.450(\underline{34}) &
41.438(\underline{34}) \\ \cline{3-5}
 & &0.000(9) & +0.016(9) & +0.027(9)  \\ \hline
 $R_l$ & 20.763(49) &
  20.749(41) & 20.742(41) & 20.738(42) \\ \cline{3-5}
   &  & 20.815(41) &
20.830(43) & 20.845(43)\\ \cline{3-5}
 &   &--0.066(\underline{5}) & --0.088(\underline{5})  & --0.107(\underline{5})
 \\ \hline
$R_b$ &
 0.2200(27) & 0.2168(6) & 0.2160(5) & 0.2154(5) \\ \cline{3-5}
  & & 0.2197(0) & 0.2197(0)& 0.2197(0) \\ \cline{3-5}
   &   & --0.0029(\underline{5})  & --0.0036(\underline{5}) &
--0.0042(\underline{5}) \\ \hline
\end{tabular}
\end{center}

\newpage

\begin{center}

{\large Table 2}

\vspace{8mm}

\begin{tabular}{|l|l|l|l|l|}\hline
\hline
 &\multicolumn{2}{l|}{LEP}&\multicolumn{2}{l|}{LEP + collider and $\nu N$
data} \\ \cline{2-5}
Parameter & ADLO \cite{ggg}& Our fit & Ref. \cite{ggg} & Our fit \\ \hline
 & & & & \\
$m_t$ (GeV) & $166^{+17 +19}_{-19 -22}$ & $162^{+16 +17}_{-17 -23}$ &
$164^{+16 +18}_{-17 -21}$ & $162^{+14 +16}_{-15 -22}$ \\
 & & & & \\
$\bar{\alpha}_s$ & $0.120(6)^{+2}_{-2}$ & $0.119(6)^{+2}_{-2}$ &
$0.120(6)^{+2}_{-2}$ & $0.119(6)^{+2}_{-2}$ \\
 & & & & \\
\hline
$\chi^2/\mbox{\rm d.o.f.}$ & 3.5/8 & 3.5/8 & 4.4/11 & 3.5/10 \\
 & & & & \\
\hline
 & & & & \\
$\sin^2\theta^{lept}_{eff}$ & $0.2324(5)^{+1}_{-2}$ & $0.2324(6)^{+1}_{-2}$
& $0.2325(5)^{+1}_{-2}$ & $0.2324(5)^{+1}_{-2}$ \\
 & & & & \\
$\sin^2\theta_W$ & $0.2255(19)^{-3}_{+5}$ & $0.2258(19)^{-3}_{+5}$
& $0.2257(17)^{-3}_{+4}$ & $0.2258(16)^{-2}_{+5}$ \\
 & & & & \\
$m_W$ (GeV) & $80.25(10)^{+2}_{-3}$ & $80.23(10)^{+2}_{-3}$
& $80.24(9)^{+1}_{-2}$ & $80.23(8)^{+1}_{-2}$ \\
 & & & & \\
\hline
\end{tabular}
\end{center}

\newpage

\begin{center}
{\bf Figure and Table Captions}
\end{center}

Fig. 1: Allowed region of $m_t$ and $\bar\alpha_s$ with $m_H$ = 300 GeV.
The lines represent the $s$-standard deviation ellipsoids ($s$=1, 2, 3, 4, 5),
corresponding to the constant values of $\chi^2$
 ($\chi^2 = \chi^2_{min} + s^2$).\\

Fig. 2: Allowed region of $m_t$ and $m_H$ with $\bar\alpha_s$ = 0.119.
The lines represent the $s$-standard deviation "ellipsoids"
 ($s$=1, 2, 3, 4, 5),
corresponding to the constant values of $\chi^2$
 ($\chi^2 = \chi^2_{min} + s^2$).\\

Fig. 3: The fitted values of $m_t$ from the individual observables measured
at LEP and $p\bar{p}$ colliders, assuming $m_H = 300$ GeV and
$\bar{\alpha}_s = 0.118$. The hatched region corresponds to $m_t < m_Z$,
which is definitely excluded by the direct search for the top quark. The
central values of $R_b$ and $R_l$ are in the excluded region.

\vspace{5mm}

Table 1: Observables: their experimental and predicted values. The first
column contains observables (four of them -- $g_A \, , \, g_V/g_A \, , \,
\Gamma_l \, , \, \Gamma_h$ -- are secondary ones connected by well-known
phenomenological relations with the observables of Fig. 3).  The second
column contains the experimental value of the observables.
Columns 3, 4, 5 have three lines for each observable: the first line
contains the predicted values which are the sum of the
$\bar{\alpha}$ Born and one-loop contributions for
$m_H = 60 \, , \, 300$, and $1000$ GeV.
The second line contains the $\bar{\alpha}$ Born values, while the third
gives one-loop contributions. The values in columns 3, 4, 5 are calculated for
the fitted values of $m_t\pm \delta m_t$ and $\bar{\alpha}_s
\pm\delta\bar{\alpha}_s$, for each value of $m_H$ respectively. Figures for
uncertainties are underlined when the coefficient in front of
$\delta\bar{\alpha}_s$ or $\delta m_t$ is negative.

Table 2: Comparison of our fit with that of the LEP Collaborations and the
LEP Electroweak Working Group (see Table 24 of ref. \cite{ggg});
$\sin^2\theta^{lept}_{eff}\equiv\frac{1}{4}(1-g_V/g_A)$,
$s^2_W\equiv\sin^2\theta_W\equiv 1-m^2_W/m^2_Z$. The collider and $\nu N$
data referred to in columns 4 and 5 were obtained in the collider
experiments UA2 \cite{hh} and CDF \cite{jj} (see Eq. (20)) and the $\nu
N$ experiments CDHS \cite{QQ}, CHARM \cite{rr} and CCFR \cite{CCFR}. The latter
been fitted (see ref.  \cite{ggg}) by $s_W^2 = 0.2256(47)$ ($m_W/m_Z =
0.8800(27)$; $m_W = 80.24(25)$ GeV). The central values in the table
correspond to $m_H = 300$ GeV, the upper and lower shifts to $m_H = 1000$
GeV and 60 GeV respectively.

\begin{figure*}[htb]\centering
\epsfig{figure=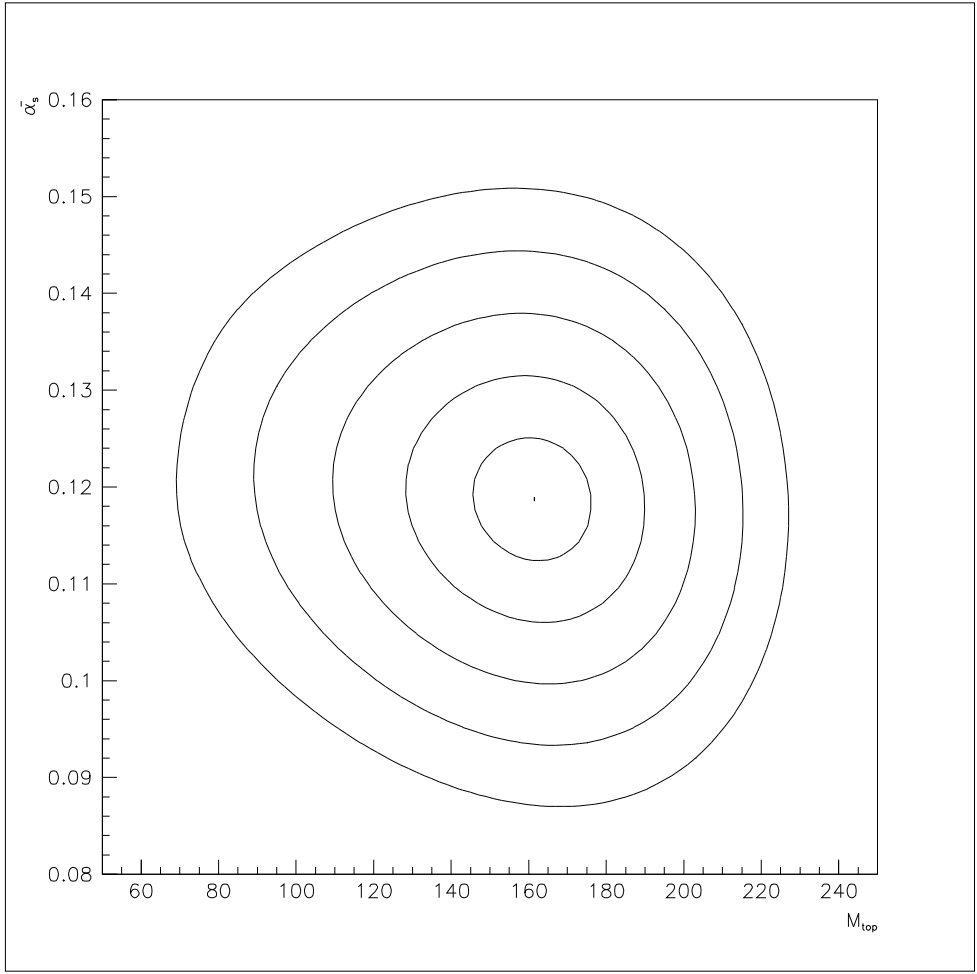,width=0.94\textwidth}
\caption[]{\label{fig1}
\rm
 Allowed region of $m_t$ and $\bar\alpha_s$ with $m_H$ = 300 GeV.
The lines represent the s-standard deviation ellipsoids
($s$=1, 2, 3, 4, 5),
corresponding to the constant values of $\chi^2$
 ($\chi^2 = \chi^2_{min} + s^2$).
}
\end{figure*}

\begin{figure*}[htb]\centering
\epsfig{figure=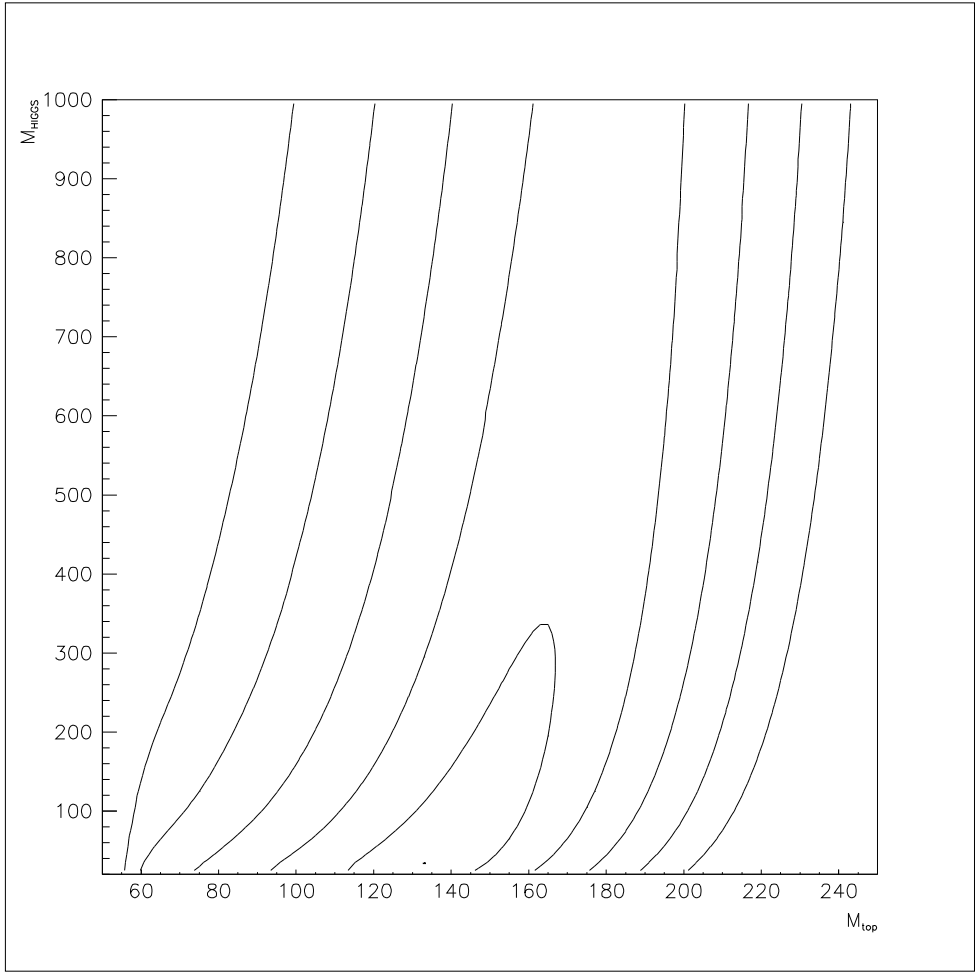,width=0.94\textwidth}
\caption[]{\label{fig2}
\rm
Allowed region of $m_t$ and $m_H$ with $\bar\alpha_s$ = 0.119.
The lines represent the s-standard deviation "ellipsoids"
 ($s$=1, 2, 3, 4, 5),
corresponding to the constant values of $\chi^2$
 ($\chi^2 = \chi^2_{min} + s^2$).
}
\end{figure*}

\begin{figure*}[htb]\centering
\mbox{\epsfig{figure=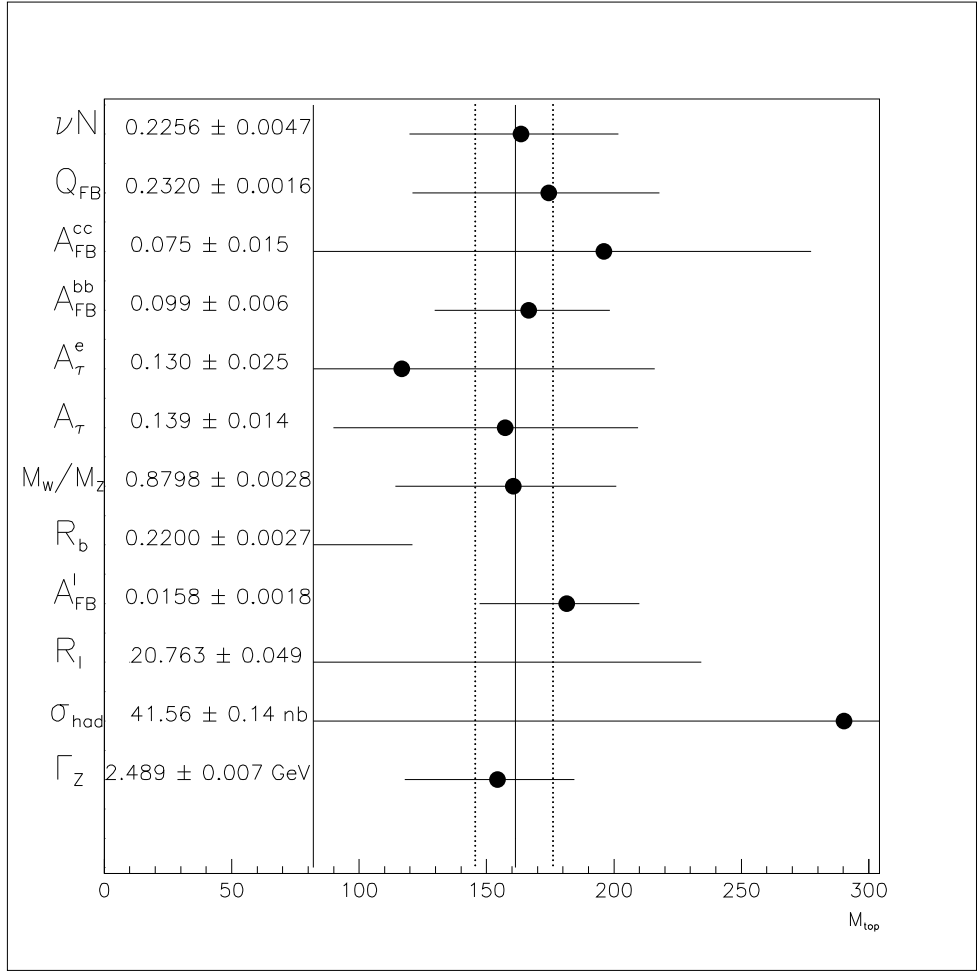,width=0.94\textwidth}}
\caption[]{\label{fig3}
\rm
Value for $m_t$ from different experimental
measurements. $\bar\alpha_s$ was fixed to 0.119.
}
\end{figure*}

\end{document}